\documentclass[letterpaper, 10 pt, conference]{ieeeconf}  
\IEEEoverridecommandlockouts                              

\usepackage{times} 
\usepackage{amsmath} 
\usepackage{amssymb}  
\usepackage[export]{adjustbox}
\usepackage[]{svg}
\usepackage[ruled,linesnumbered]{algorithm2e}
\DeclareMathAlphabet{\mathpzc}{OT1}{pzc}{m}{it}
\usepackage{graphicx}
\usepackage{multirow}
\usepackage{pifont}%
\newcommand{\cmark}{\ding{51}}%
\newcommand{\BibTeX}{\rm B\kern-.05em{\sc i\kern-.025em b}\kern-.08em\TeX}
\usepackage{booktabs}
\newtheorem{definition}{Definition}
\newtheorem{proposition}{Proposition}
\newtheorem{theorem}{Theorem}
\usepackage{ifthen}

\usepackage{etoolbox}
\providetoggle{beginFlag}
\settoggle{beginFlag}{true}

\usepackage{cite}

\title{\LARGE \bf
BRNES: Enabling Security and Privacy-aware Experience Sharing in Multiagent Robotic and Autonomous Systems
}

\author{Md Tamjid Hossain, Hung Manh La, Shahriar Badsha, and Anton Netchaev
\thanks{Md Tamjid Hossain and Hung La are with the Advanced Robotics and Automation Lab, Computer Science and Engineering,
        University of Nevada, Reno, Nevada, USA.
        {\tt\small Emails: mdtamjidh@nevada.unr.edu,  hla@unr.edu}}%
\thanks{Shahriar Badsha is with Bosch Engineering, North America.
        {\tt\small Email: shahriar.badsha@us.bosch.com}}
\thanks{Anton Netchaev is with the U.S. Army Corps.
        {\tt\small Email: anton.netchaev@erdc.dren.mil}}        
\thanks{This work was partially funded by the U.S. National Science Foundation (NSF) under grants: NSF-CAREER: 1846513, and  NSF-PFI-TT: 1919127.  The views, opinions, findings, and conclusions reflected in this publication are solely those of the authors and do not represent the official policy or position of the NSF.}
\thanks{*Source code of the task and the computational model behind the setup available at {\tt \small https://github.com/aralab-unr/BRNES}}%
}

\begin{document}

\maketitle
\thispagestyle{empty}
\pagestyle{empty}

\begin{abstract}
Although experience sharing (ES) accelerates multiagent reinforcement learning (MARL) in an advisor-advisee framework, attempts to apply ES to decentralized multiagent systems have so far relied on trusted environments and overlooked the possibility of adversarial manipulation and inference. Nevertheless, in a real-world setting, some Byzantine attackers, disguised as advisors, may provide false advice to the advisee and catastrophically degrade the overall learning performance. Also, an inference attacker, disguised as an advisee, may conduct several queries to infer the advisors' private information and make the entire ES process questionable in terms of privacy leakage. To address and tackle these issues, we propose a novel MARL framework (BRNES) that heuristically selects a dynamic neighbor zone for each advisee at each learning step and adopts a weighted experience aggregation technique to reduce Byzantine attack impact. Furthermore, to keep the agent's private information safe from adversarial inference attacks, we leverage the local differential privacy (LDP)-induced noise during the ES process. Our experiments show that our framework outperforms the state-of-the-art in terms of the steps to goal, obtained reward, and time to goal metrics. Particularly, our evaluation shows that the proposed framework is 8.32x faster than the current non-private frameworks and 1.41x faster than the private frameworks in an adversarial setting.

\end{abstract}


\section{INTRODUCTION}
\label{intro}
Experience sharing (ES) \cite{da2017simultaneously} has become increasingly significant in the multiagent reinforcement learning (MARL) \cite{la2014multirobot} paradigm due to its efficacy in accelerating learning performance. As the popularity of ES processes increases, so do concerns about their security and privacy. Namely, advisors' shared experience shapes the learning behavior and outcomes of an advisee \cite{da2017simultaneously}. A shared but malicious experience could mislead an advisee to take incorrect measures during the \textit{experience harvesting (EH)} phase of ES \cite{ye2020differentially,figura2021adversarial}. Likewise, as the shared experience is computed based on the inputs (e.g., reward signal) that commonly rely on advisors' data, an inference attack on those may leak advisors' private information during the \textit{experience giving (EG)} phase of ES \cite{wang2019privacy, prakash2022private}. These security (adversarial manipulation) and privacy (adversarial inference) threats, unfortunately, overlooked by many related studies \cite{da2017simultaneously,mahdavimoghadam2022improved, li2020cooperative,hussein2022autonomous,matta2019q}; can bring down catastrophic consequences on MARL-based safety-critical applications in domains such as robotics \cite{la2014multirobot}, cyber-physical systems \cite{prasad2019multi}, automotive industries \cite{kiran2021deep}, etc. For example, false advising from an advisor car in autonomous driving may make lane-changing ambiguous and lead to severe road accidents for an advisee car \cite{Zhou2022}, whereas an inference attack from an advisee may reveal sensitive data of the advisors \cite{wang2019privacy, Farhan2023}.
\textit{Therefore, to facilitate a secure and private MARL for next-generation robotic and autonomous systems, a study of the adversarial manipulation and inference threats posed by the current ES process is non-trivial.}

Particularly, from a security perspective, false advising
threat is prominent in the decentralized MARL settings, where there is no central authority to ensure the consensus on advice integrity and agents' authenticity, and thus susceptible to Byzantine general problems \cite{lamport2019byzantine}. Researchers in 
\cite{ye2020differentially} address this false advising threat from Byzantine advisors in a MARL platform by adopting differential privacy (DP) \cite{dwork2006} at the advisee's end. However, a strategic attacker can exploit the DP-noise to conduct optimal false data injection (or simply false advising) attacks and hamper the learning outcomes significantly \cite{giraldo2020adversarial,hossain2021privacy}. \textit{To tackle this, we propose to incorporate the experience, whether it is differentially private or not, into the advisee's learning through a weighted experience aggregation technique.} 

From a privacy perspective, we argue that inference attackers, disguised as advisees, could try to infer advisors' sensitive information by recursively querying their experience for every state-action pair. For example, the advisors' experience in Q-value sharing frameworks (e.g., \cite{zhu2021q,da2017simultaneously,ye2022differential}) can reflect their \textit{rewarding strategy} that builds their decision-making criteria, and \textit{movement trajectory} that carries important contextual information, such as users' preference, next course of actions, etc. 
\cite{wang2019privacy}. \textit{To protect such sensitive information in untrusted environments, unlike \cite{ye2020differentially}, we propose to adopt local differential privacy (LDP) \cite{Kasiviswanathan2008}} during ES. LDP perturbs advisors' experience before sharing it, making it harder for inference attackers to obtain sensitive information.
Different from the above-mentioned works, our paper presents a novel Byzantine Robust Neighbor Experience Sharing (BRNES) framework that addresses the security and privacy threats in the ES process from two adversarial perspectives: (1) false advising during the advisee's EH, and (2) inference attack during the advisor's EG. Therefore, our contribution in this paper is twofold: in decentralized EH, we address the security attacks from \textit{Byzantine attackers}, and in EG, we address the privacy attacks from inference attackers. To ensure the framework is Byzantine robust, we develop an \textit{adaptive heuristic neighbor zone selection} process for each advisee that limits the possibility of a Byzantine advisor deterministically appearing in the vicinity of any targeted advisee significantly due to the inherent randomness in the process. Additionally, to further limit the false advising impacts from Byzantine advisors, we leverage a \textit{weighted experience aggregation} technique that prevents the direct integration of advisors' experience. To prevent inference attackers from inferring advisors' sensitive data, we leverage the provable privacy guarantee offered by the LDP mechanism.
In summary, our contributions are:
\begin{itemize}
    \item 
    to enable security and privacy-aware ES in MARL, we propose a novel framework (BRNES) that addresses adversarial manipulation and inference problems in existing multiagent robotic and autonomous systems and fills two important research gaps in the literature- (1)\textit{ the absence of a Byzantine robust decentralized EH mechanism}, and (2)\textit{ the lack of a private EG process}.
    
    \item to achieve Byzantine robustness, we formulate a novel adaptive heuristic neighbor zone selection strategy and leverage the weighted experience aggregation technique. 
    \item to make EG privacy-protected, we leverage the provable privacy guarantee offered by the LDP technique. 
    \item comparing to the state-of-the-art (SOTA), we show that our framework is $1.41$x faster than DA-RL \cite{ye2022differential} and $8.32$x faster than AdhocTD \cite{da2017simultaneously} under adversarial presence.
\end{itemize}

\section{Related Works}

ES strategies have been studied extensively to enhance the learning performance of MARL agents \cite{da2017simultaneously, mahdavimoghadam2022improved, ye2020differentially, ye2022differential, cheng2021general, da2018autonomously, figura2021adversarial,le2017coordinated, li2020cooperative, matta2019q, nguyen2019hindsight,rashid2018qmix,zhu2021q}. For instance, the problem of slow convergence of MARL policy is addressed in \cite{mahdavimoghadam2022improved}, where, to tackle the slow learning, the authors propose central knowledge transfer units for the participating agents. 
Similarly, to mitigate the curse of dimensionality in conventional ES-driven MARL platforms, several novel MARL algorithms based on mixed Q-networks \cite{rashid2018qmix}, simultaneous learning \cite{matta2019q, da2017simultaneously}, and differential advising
\cite{ye2020differentially,ye2022differential} have been developed.


Specifically, \cite{da2017simultaneously} reduces the number of inter-agent communications by (1) limiting the students to seek advice from the teachers only when their confidence is low for a given state, and (2) limiting the teachers to respond only when they believe they have much knowledge for that state. \textit{Nonetheless, \cite{da2017simultaneously} overlooks the possibility of adversarial manipulation and inference, which may impede the success of ES processes in real-life MARL applications.}

An alternative approach to simultaneous learning, the iteration-based Q-learning is proposed in \cite{matta2019q}, where a \textit{centralized aggregator} forms a swarm matrix containing the extremes of Q-values from all agents. 
\textit{Nonetheless, \textit{centralized aggregation} may possess various drawbacks (e.g., single-point-of-failure) despite its fast convergence.}

Intuitively, a decentralized mechanism is more effective in an environment with resource-constrained edge devices 
than a centralized mechanism. Moreover, decentralization alleviates the single-point-of-failure problem. Considering this, \cite{zhu2021q} introduces a decentralized and heuristic Q-value advising method called PSAF that addresses \textit{when to ask for the advice, when to give the advice, and how to use the advice} in a teacher-student framework. \textit{However, decentralization may create opportunities for the Byzantine and inference attackers \cite{lamport2019byzantine}.} 
Field research and experience of MARL application's post-deployment \cite{figura2021adversarial,cheng2021general} show that any malicious agent, in general, may conduct eavesdropping, inference attacks, Byzantine attacks, etc., creating significant security and privacy challenges for multiagent systems (MAS).

Researchers partially solve the false advising in MARL \cite{ye2020differentially}. They design the adviser selection problem as a Multi-armed bandit and solve it using the DP technique. However, their assumption of eliminating probabilistic false advice by malicious agents through direct DP integration does not hold in the presence of any strategic attacker. The extension of their work involves accommodating the advice from a slightly different state \cite{ye2022differential}. 
\textit{Yet, they adopt the DP mechanism for learning performance improvement only, but not to protect the privacy and security of the agents.}

Privacy and security concerns in MAS are addressed in \cite{figura2021adversarial, li2022privacy}. From the privacy perspective, \cite{li2022privacy} emphasizes preserving agents' privacy against inference attackers by proposing a DP-MAS framework. From the security perspective, \cite{figura2021adversarial} shows that an adversary can mislead honest agents to attain its malicious objectives in a consensus-based MARL platform. \textit{However, both \cite{figura2021adversarial, li2022privacy} are limited to centralized environments, and thus, cannot apply to decentralized MARL applications.} We summarize major contrasting points between literature and this work in Table~\ref{tab:litcomparison}.

\begin{table}[!ht]
\centering
\caption{MARL framework comparison. Symbol: Addressed ($\textbf{\cmark}$), Not addressed ($\square$).
``L"earning type (``C"entral or ``D"ecentral). ``H"euristic advising . ``A"dvising confidence. ``B"udget constraints. ``F"alse advising. ``P"rivacy attacks. ``N"eighbor zone. ``\textbf{W}"eighted advice aggregation.}
\resizebox{\linewidth}{!}{%
\begin{tabular}{lccccccccc} 
\toprule
                    & \multicolumn{2}{c}{L} & \multirow{2}{*}{H} & \multirow{2}{*}{A} & \multirow{2}{*}{B} & \multirow{2}{*}{F} & \multirow{2}{*}{P} & \multirow{2}{*}{N} & \multirow{2}{*}{W}  \\ 
\cline{2-3}
                    & C    & D              &                    &                    &                    &                    &                    &                    &                     \\ 
\midrule
Silva et al., 2017 \cite{da2017simultaneously}  & $\square$  & \textbf{\cmark}           & \textbf{\cmark}               & \textbf{\cmark}               & \textbf{\cmark}               & $\square$                & $\square$                & $\square$                & $\square$                 \\
Matta et al., 2019 \cite{matta2019q}  & \textbf{\cmark} & $\square$            & $\square$                & $\square$                & $\square$                & $\square$                & $\square$                & $\square$                & \textbf{\cmark}                \\
Ye et al., 2020 \cite{ye2020differentially}    & $\square$  & \textbf{\cmark}           & \textbf{\cmark}               & \textbf{\cmark}               & \textbf{\cmark}               & \textbf{\cmark}               & $\square$                & $\square$                & $\square$                 \\
Figura et al., 2021 \cite{figura2021adversarial} & \textbf{\cmark} & $\square$            & $\square$                & $\square$                & $\square$                & \textbf{\cmark}               & \textbf{\cmark}               & $\square$                & \textbf{\cmark}                \\
Zhu et al., 2021 \cite{zhu2021q}   & $\square$  & \textbf{\cmark}           & \textbf{\cmark}               & \textbf{\cmark}               & $\square$                & $\square$                & $\square$                & $\square$                & $\square$                 \\
Li et al., 2021 \cite{li2022privacy}    & \textbf{\cmark} & $\square$            & $\square$                & $\square$                & $\square$                & $\square$                & \textbf{\cmark}               & $\square$                & \textbf{\cmark}                \\
Ye et al., 2022 \cite{ye2022differential}    & $\square$  & \textbf{\cmark}           & \textbf{\cmark}               & \textbf{\cmark}               & \textbf{\cmark}               & $\square$                & $\square$                & $\square$                & $\square$                 \\
\textbf{This work}           & $\square$  & \textbf{\cmark}           & \textbf{\cmark}               & \textbf{\cmark}               & \textbf{\cmark}               & \textbf{\cmark}               & \textbf{\cmark}               & \textbf{\cmark}               & \textbf{\cmark}                \\
\bottomrule
\end{tabular}
}
\label{tab:litcomparison}
\end{table}
\section{Problem Formulation and Threat Modelling}
Let us consider $\mathpzc{N}$ robotic agents ($\mathpzc{N} = \{p_1, ..., p_n\}$), which are learning cooperatively to achieve an objective in environment $\mathbb{E}$ of $\mathpzc{H} \times \mathpzc{W}$ dimension following a Markov game. The game is represented as a tuple $(\mathpzc{N}, \mathpzc{S}, \mathpzc{A}, \Phi, \gamma, \mathpzc{T})$ having state-space $\mathpzc{S} = \mathpzc{S}_1 \times ... \times \mathpzc{S}_n$, joint action space $\mathpzc{A} := \mathpzc{A}_1 \times \mathpzc{A}_2 \times ... \times \mathpzc{A}_n$, transition function $\mathpzc{T}: \mathpzc{S} \times \mathpzc{A}$, reward function $\Phi: \mathpzc{S} \times \mathpzc{A} \times \mathpzc{S}$, and discount factor $\gamma \in \left[0, 1\right]$ for all future rewards. 
The goal point is $\mathpzc{G}$ and the action space is $\mathpzc{A} = \{Left, Right, Up, Down\}$. Consider several obstacles $\mathpzc{O}_{x = \{0, 1, ...\}}$ in the environment. If any agent hits the environment boundaries or the obstacles, it would get a penalty, $\phi_{\mathpzc{O}}\in \Phi$. However, assume one freeway $\mathpzc{F}$ in $\mathbb{E}$, which could be used by any agent to earn a reward before reaching $\mathpzc{G}$. Incorporating this freeway structure within the grid-based environmental model enhances the opportunities for agents to accrue supplementary rewards, and invariably introduces additional dimensions of complexity to the task landscape. 
After reaching $\mathpzc{G}$, the agents are rewarded with $\phi_{\mathpzc{G}}\in \Phi$ s.t. $\phi_{\mathpzc{G}} >\phi_{\mathpzc{F}}$ (if $\phi_{\mathpzc{G}} \leq \phi_{\mathpzc{F}}$, the agents would not be motivated to move to the goal). Note that, $\lvert \mathpzc{N}\rvert + \lvert \mathpzc{O_x}\rvert + \lvert \mathpzc{F}\rvert + \lvert \mathpzc{G}\rvert < \mathpzc{H}\times \mathpzc{W}$, otherwise the agents cannot move smoothly through the empty spaces of the grid. 

The position of the agents, obstacles, and freeway are randomly initialized in each episode. 
Assume the obstacles randomly change positions at each step, thus making it harder for the agents to learn. The objective completes when all the agents reach $\mathpzc{G}$. Any individual agent $p_i$ is spatially aware of the position of $\mathpzc{G}$. \textbf{$p_i$'s objective is to take the lowest possible steps to goal ($SG_{min}$) for collecting freeway reward $\phi_f$ and reaching goal $\mathpzc{G}$ without hitting environment boundary or any obstacles $\mathpzc{O}_x$ while, also, earning maximum rewards ($\phi_{max} = \phi_{\mathpzc{F}} + \phi_{\mathpzc{G}} + [\phi_{\mathpzc{O}} = 0]$)}. Thus, $p_i$'s objective can be formalized as (a) $SG_{p_i} = SG_{min}$, (b) $\phi_{p_i} = \phi_{max}$, and (c) $\lVert (x_{p_i}, y_{p_i})- (x_{\mathpzc{G}}, y_{\mathpzc{G}}) \rVert = 0$,  
where $(x_{p_i}, y_{p_i})  \in [(0,0), (\mathpzc{H}\times \mathpzc{W})]$ and  $(x_{\mathpzc{G}}, y_{\mathpzc{G}}) \in [(0,0), (\mathpzc{H}\times \mathpzc{W})]$ are $p_i$'s and $\mathpzc{G}$'s position,  respectively.

\subsection{Byzantine Attacks during EH} During EH, a Byzantine advisor $p_b \in \mathpzc{N}$ may send false information to $p_i$, with a
\textit{malicious objective to impede $p_i$'s convergence} as depicted in Figure~\ref{fig:threatModel}a. We assume that $p_b$ has the knowledge of $\mathpzc{A, S}, \Phi, (x_{\mathpzc{G}},y_{\mathpzc{G}})$ and $p_{i}$'s current state $s$. Particularly, $p_b$ could promote a larger $Q$-value for a misleading action $a_m \in A$ than the rest of the actions $\mathpzc{A}$\textbackslash$\{a_{m}\}$, i.e., $Q_{p_b}(s^t, a_{m}) > Q_{p_b} (s^t, a_h) \forall a_h \in \mathpzc{A}$\textbackslash$\{a_m\}$, thus continuously drive $p_i$ towards a desired malicious point. However, if $p_b$ always shares a set of large Q-values to attain a large incentive, it might be identified easily by any anomaly detector at the advisee's end. On the contrary, if it shares a set of small Q-values, the attack impact might be negligible. This fundamental adversarial tradeoff problem can be tackled in several ways. One approach is to shuffle the Q-values for all actions corresponding to the requested state and inject false noise that is similar to the maximum reward using reward poisoning methods \cite{zhang2020adaptive}. Another approach is to draw the false noise from an adversarial distribution that has similar statistical properties to a benign noise distribution used for achieving DP \cite{giraldo2020adversarial}. For simplicity, but without losing generality, we choose the former method to generate false advice in this study since any optimal false advising attack method would always involve false data that is difficult to distinguish from benign data.
    

\subsection{Inference Attacker during EG} 
\begin{figure}[!t]
\centerline{\includegraphics[width=\linewidth]{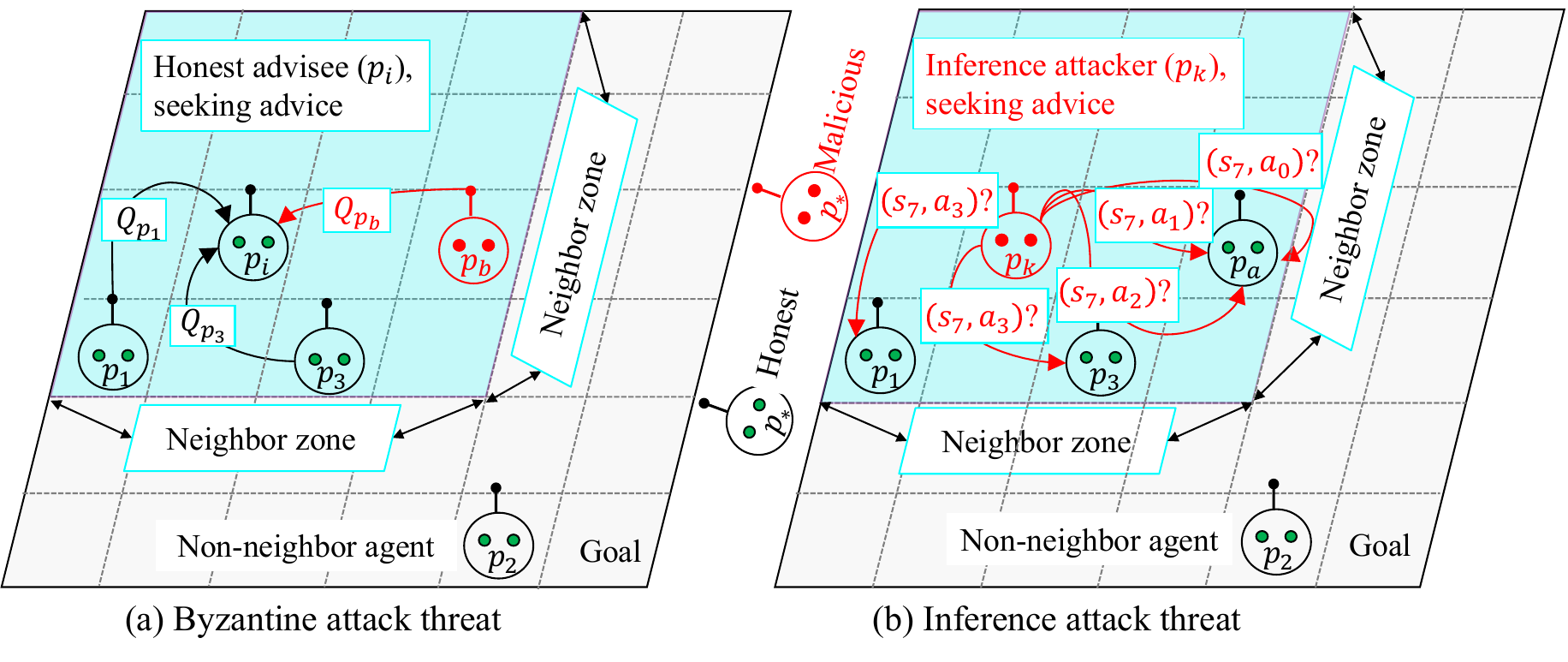}}
    \caption{Threats in MARL: (a) A Byzantine advisor ($p_b$) providing false information ($Q_{p_b}$) to the honest advisee ($p_i$); (b) An inference attacker ($p_k$) performing multiple queries ($s_7, a_{0,1,...}$) to an advisor ($p_a$).} 
    \label{fig:threatModel}
    \vspace{-15pt}
\end{figure}

The advisee itself could be an adversary, whose \textit{malicious objective is to infer the private information of an honest advisor $p_a \in \mathpzc{N}$ by analyzing $p_a$'s experience} (Figure~\ref{fig:threatModel}b). We assume that the advisee has the knowledge of $\mathpzc{A, S}, \Phi, (x_{\mathpzc{G},y_{\mathpzc{G}}})$, but does not know $p_{a}$'s current state. Specifically, a malicious advisee $p_k \in \mathpzc{N}$ could perform multiple queries to $p_a$'s Q-tables for each and every state and action in order to reconstruct $p_a$'s entire Q-table and infer sensitive information related to $p_a$'s residing states, next actions, rewards, and adopted strategies. 

\section{BRNES Framework}
\begin{figure}[!b]
\centering
\vspace{-5pt}
    \includegraphics[width=\columnwidth]{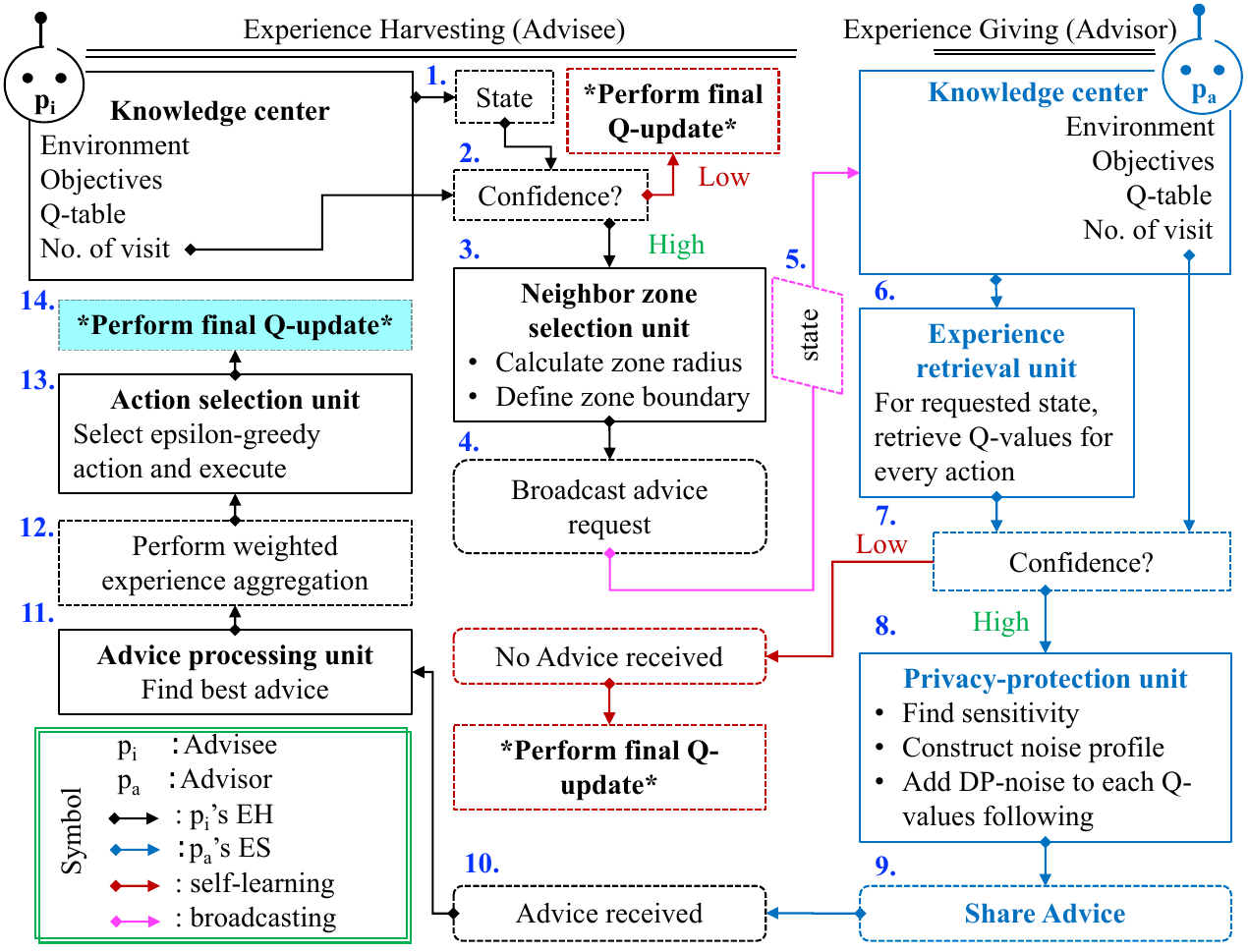}
    \caption{BRNES framework: Advisee $p_i$ is harvesting the experience while advisor $p_a$ is sharing experience to $p_i$.} 
    \label{fig:BernetFramework}
\end{figure}

We model our BRNES framework (Figure ~\ref{fig:BernetFramework}) for robotic agents which share their experience under the adversarial presence and budget constraints. We use a model-free and off-policy MARL approach, Q-learning, to develop and test our framework in a stochastic environment. We formulate the experience as Q-values instead of the recommended actions since the Q-value advising, unlike the action advising, does not impair the performance of the agent's learning directly \cite{zhu2021q}. The framework mimics an advisee-advisor network where agents are homogeneous and interchangeable. They have identical strategies, however, they maintain their own Q-table to store their local knowledge. 
Algorithm~\ref{algo1} presents the pseudocode for the EH phase. Algorithm~\ref{algo2} outlines the required sub-functions and Algorithm~\ref{algo3} shows the LDP adaptation technique during the EG phase.

\SetAlgoSkip{SkipBeforeAndAfter}
\begin{algorithm}[!t]

\SetKwInput{KwNotation}{Notation}
\SetKwInput{KwInput}{Require} 
\DontPrintSemicolon
\KwInput{Environment, $\mathbb{E}$}
\textbf{Initialize} Q-table
and \textbf{set} $\epsilon, \alpha, \gamma$\\
\For{each $t = 1, 2, ..., T$ episodes}
{
Observe $s^t_{p_i}$, find $n_{p_i}^{v} \gets s^t_{p_i}$, and 
compute $P_{p_i}^a= f(n_{p_i}^{v}, B_{p_i}, B_{p_i}^{tot}, \tau, \tau'$) from \textit{Algorithm $2$}\\

\eIf{$0<P_{p_i}^a<\kappa$}  
{
Find $\mathpzc{Z}_{p_i}^t$ = $NZ$($\lvert \mathpzc{N}\rvert, s_{p_i}^t, \mathbb{E}$) from \textit{Algorithm $2$}\\
Send advise request to neighbors within $\mathpzc{Z}_{p_i}^t$\\ 
\eIf{No Advice}{
Perform final Q-update
}
{Receive advice from all $k$ advisors as $\left[Q'_{p_a}(s_{p_i}^t)\right]_{p_a = 1}^k$ (refer to \textit{Algorithm 3})\\
Find best advice ($\xi_{p_i}^t$) by grouping \& averaging Q-values for each action\\ 
Perform weighted aggregation
$Q_{p_i}(s_{p_i}^t)= w \times Q_{p_i}(s_{p_i}^t) + (1-w)\times \xi^{t}_{p_i}$\\
Find $\epsilon$-greedy action \& observe $s_{p_i}^{t+1}$, $\Phi_{p_i}^t$\\
Perform final Q-update for selected action
$Q_{p_i}^t (s_{p_i}^t, a^t) = (1-\alpha)Q_i^t(s_{p_i}^t, a^t) + \alpha(\Phi_{p_i}^t + \gamma * \underset{a^{t+1}}{max}\;Q_{p_i}^t(s_{p_i}^{t+1}, a^{t+1}))$\\
}
}
{
Take action and perform final Q-update
}
Set $s_{p_i}^t = s_{p_i}^{t+1}$ \tcp*[f]{update to next state}\\
\textbf{if} $\lVert (x_{p_i}, y_{p_i})- (x_{\mathpzc{G}}, y_{\mathpzc{G}}) \rVert > 0$ \textbf{then} Continue\\ \textbf{else} End episode and reset environment
}
\caption{Experience harvesting (EH) by advisee $p_i$. $\mathpzc{G}$: goal, $p_i$: advisee, $\mathbb{E}$: environment, $n^{v}$: number of visit, $\epsilon$: probability, $\alpha$: learning rate, $s$: state, $a$: action, $\Phi$: reward set, $\gamma$: discount factor, $\mathpzc{Z}$: neighbor zone, $\mathpzc{N}$: agent set, $\xi$: best advice, $(x,y)$: position coordinate, $w$: aggregation factor (weight), $B$: advice seeking budget, $\tau,\tau', \kappa$: predefined threshold}

\label{algo1}
\end{algorithm}
\vspace{-3pt}
\subsection{Experience Harvesting (EH) Process}
\vspace{-3pt}
To tackle the adversarial manipulation, it is necessary to ensure that no particular advisor frequently appears in the close vicinity of advisee $p_i$ for multiple episodes and that their advice is not directly integrated into $p_i$'s Q-learning. Considering this, at timestamp $t$, $p_i$ first observes its current state initialized by a stochastic initialization process (i.e., at every episode, all agents appear in random states, thus limiting the consecutive attack opportunity over a targeted agent) (Figure ~\ref{fig:BernetFramework}, step $1$). Then, $p_i$ computes its experience harvesting confidence (EHC), $P_{p_i}^a$ (Algorithm~\ref{algo1}, line $3$). The advisee seeks advice from the experienced advisors in its neighborhood only when- (1) its knowledge of that particular state is low, and (2) it has the budget to seek advice. 

\subsubsection{Computing Experience Harvesting Confidence (EHC)} $p_i$'s EHC can be calculated as Algorithm~\ref{algo2}, line $1-2$ \cite{ye2022differential}, where $p_i$'s current and total communication budget are $B_{p_i}$ and $B_{p_i}^{tot}$, respectively. The user-defined threshold $\tau$ prevents $p_i$ to avoid spending all of its budgets in the early episodes and $\tau'$ prevents $p_i$ to avoid seeking advice for the highly-visited states. Function, $f$ provides a higher probability for the states that the advisee visits rarely and vice versa. $p_i$ performs final Q-update if $P_{p_i}^a$ is zero. Otherwise (i.e., $0<P_{p_i}^a<\kappa$ where $\kappa$ is a predefined threshold), it proceeds to the next steps as shown in line $5-14$ of Algorithm~\ref{algo1}.

\begin{algorithm}[!t]
\SetKwFunction{FConf}{$f$}
\SetKwFunction{FNeigh}{$NZ$}
\DontPrintSemicolon
\SetKwProg{Fn}{Function}{:}{}
\Fn{\FConf{$n_{p_i}^{v}, B_{p_i}, B_{p_i}^{tot}, \tau, \tau'$}}
{
\KwRet $P_{p_i}^a = $
$\begin{cases}
    \frac{1}{\sqrt{n_{p_i}^{v}}} \cdot \sqrt{\frac{B_{p_i}}{B_{p_i}^{tot}}}, & \tau \leq n_{p_i}^{v} \leq \tau'\\
    0, & Otherwise
\end{cases}$

}
\SetKwProg{Fn}{Function}{:}{}
\Fn{\FNeigh{$\lvert \mathpzc{N} \rvert, s_{p_i}^t, \mathbb{E}$}}
{
\textbf{Find} $p_i$'s position at $t$, i.e., $(x_{p_i}^t, y_{p_i}^t)\gets s_{p_i}^t$\\
\textbf{Find} height($\mathpzc{H}$), width($\mathpzc{W}) \gets \mathbb{E}$\\
\textbf{Calculate} zone radius, $r_{p_i}^t = \sqrt{\frac{\mathpzc{H}\times \mathpzc{W}}{\lvert \mathpzc{N} \rvert}}$\\
\textbf{Define} zonal boundary lines, $\mathpzc{Z_{p_i}^t}=\left[x_{p_i}^t \pm r_{p_i}^t, y_{p_i}^t \pm r_{p_i}^t\right]$
$\forall 0\leq (x_{p_i}^t \pm r_{p_i}^t)\leq\mathpzc{W} \text{ and } 0\leq (y_{p_i}^t \pm r_{p_i}^t)\leq\mathpzc{H}$\\
\KwRet $\mathpzc{Z}_{p_i}^t$\;
}
\caption{Sub-functions.}
\label{algo2}
\end{algorithm}
\subsubsection{Selecting Adaptive Heuristic Neighbor Zone} \label{Neighborzone} To avoid any specific agent from frequently appearing in the neighbor zone, $p_i$ computes the radius of the neighbor zone based on the environment's dimensions and the total number of agents (Algorithm~\ref{algo2}, line $4-7$). Since we use a 2D grid space, we only consider the $x$ and $y$ coordinates of the environment when calculating the neighbor zone. However, in more complex environments with multiple dimensions, the neighbor zone to those dimensions could be extended. If there are few agents in a large grid space, the zone radius would be large, but if the agent number increases or the grid space gets smaller, the zone radius would become smaller. The boundary of the zone is calculated at each timestamp, and it is adjusted as $p_i$ moves to a new state in each episode. Since the zone size is dynamic and shifts with $p_i$'s movement, the chance of the same manipulative advisor repeatedly appearing in $p_i$'s neighbor zone is reduced.

\subsubsection{Performing Weighted Experience Aggregation} Advisee $p_i$ seeks advice from the agents residing in its neighbor zone. If no advice is received, the EH process is terminated, and the final Q-update is computed. Nonetheless, if $p_i$ receives advice, then it computes the best advice set of Q-values ($\xi_{p_i}^t$) by grouping and averaging Q-values (i.e., $\xi_{p_i}^t \leftarrow \frac{1}{n}\sum_{p_a=0}^{k}Q_{p_a}$) for every action (Algorithm \ref{algo1}, line $10-11$). After that, $p_i$ incorporates the best advice into its Q-table following a weighted linear combination process (Algorithm~\ref{algo1}, line $12$). The degree of advice is controlled by a user-defined weight factor $w\in\left[0,1\right]$. This ensures that even if any Byzantine advisor $p_b$ provides false information with the highest Q-value, it should not affect $p_i$'s learning significantly. Next, $p_i$ performs the conventional $\epsilon$-greedy action and observes the next state and reward. Finally, $p_i$ performs the final Q-update 
and update its state (Algorithm \ref{algo1}, line $13-14, \text{and } 19$).
\vspace{-2pt}
\subsection{Experience Giving (EG) Process}
When advisor $p_a$ receives an advice request from an advisee $p_i$ for any state, it has to solve the following problems: (1) whether it is confident enough to provide the advice, and (2) if it is safe to provide the advice. 
\subsubsection{Computing Experience Giving Confidence (EGC)} To tackle the first problem, we use the experience giving confidence (EGC) process described in Algorithm~\ref{algo2}, line $2$. Specifically, $p_a$ computes a probability of giving advice, $P_{p_a}^g$ based on its knowledge about that state (i.e., visit time, $n_{p_a}^v$ and advice giving budget, $B_{p_a}$. If $P_{p_a}^{g}$ is zero, $p_a$ does not provide any advice to $p_i$.

\subsubsection{Incorporating Local Differential Privacy (LDP)}\label{ES_incorporatingDP} To solve the adversarial inference problem, advisor, $p_a$ uses the DP technique that ensures that the output of an algorithm is not affected by small changes in input data from individual users. DP is typically set up in a way that involves a trusted third party, who collects data, adds noise to the query results in a way that meets the DP requirements, and then releases the noisy results. Nonetheless, in practice, finding a trusted third party could be difficult \cite{Kasiviswanathan2008}. For example, in our threat model, the advisee itself could be an untrusted party. To address this issue, the $\varepsilon$-LDP mechanism \cite{Kasiviswanathan2008}, a variant of the basic DP technique \cite{dwork2006}, emerges. $\varepsilon$-LDP applies the DP property locally to each user's data following a predefined privacy budget ($\varepsilon$) without the need for a trusted third party, rather than to the data as a whole. 
The formal definition of the $\varepsilon$-LDP mechanism can be given as \cite{Kasiviswanathan2008}:

\begin{definition}
    A randomized mechanism $\mathcal{M}$ satisfies $\varepsilon$-LDP if for any pairs of input values $x$ and $x'$ in the domain of $\mathcal{M}$, and for any possible output $y\in\mathcal{Y}$, it holds
    \begin{equation}
        \mathbb{P}\left[\mathcal{M}(x)=y\right]\leq e^{\varepsilon}\cdot\mathbb{P}\left[\mathcal{M}(x')=y\right],
    \end{equation}
    \label{def_LDP}
    \vspace{-10pt}
\end{definition}
where $\mathbb{P\left[\cdot\right]}$ denotes probability, $\mathcal{Y}$ denotes output domain, and $\varepsilon$ is the privacy budget. The smaller the $\varepsilon$, the stronger the privacy protection, but the weaker the data utility, and vice versa. $\varepsilon$-LDP allows advisors to have \textit{plausible deniability} whether or not the advisee is compromised. It satisfies the sequential property that facilitates the development of complex LDP algorithms from simpler subroutines and can be described as \cite{Kasiviswanathan2008}: 
\begin{theorem}
If $\mathcal{M}_i(x)$ is an $\varepsilon_i$-LDP algorithm for $x$ and $\mathcal{M}(x)$ is the sequential composition of $\mathcal{M}_1(x), ..., \mathcal{M}_n(x)$, then $\mathcal{M}(x)$ satisfies $\varepsilon$-LDP for $\varepsilon= \sum_{i=1}^{n}\varepsilon_i$.
\end{theorem}

Further details and the proof of Theorem $1$ can be found in \cite{Kasiviswanathan2008}. 
\begin{algorithm}[!t]
\SetKwInput{KwNotation}{Notation}
\SetKwInput{KwInput}{Require}
\DontPrintSemicolon
\KwInput{ $s_{p_i}^t, n^{v}_{p_a}, \varepsilon$}
\textbf{Receive} advice request for state $s_{p_i}^t$ from advisee $p_i$\\
$P_{p_a}^{g}$ =
$\begin{cases}
    1 - \frac{1}{\sqrt{n^{v}_{p_a}}} \cdot \sqrt{\frac{B_{p_a}}{B_{p_a}^{tot}}}\text{,} & n^{v}_{p_a} > n^{v}_{p_i}\\
    0\text{,} & \text{Otherwise}
\end{cases}$\\
\If{$P^{g}_{p_a}>0$}
{\For{each Q-value, $x$ in set $Q_{p_a}(s_{p_i}^t)$}
{$b$ = random.random()\\
\textbf{if }{$b\leq e^{\frac{\varepsilon}{n}}/(d+e^{\frac{\varepsilon}{n}}-1)$} \textbf{then}
{$Q'_{p_a}(s_{p_i}^t) \gets x$}\\
\textbf{else} 
{$Q'_{p_a}(s_{p_i}^t) \gets $\text{ Uniform($Q_{p_a}(s_{p_i}^t)/x$)}
}
}
\KwRet $Q'_{p_a}(s_{p_i}^t)$
} \textbf{else }\KwRet $\text{No Advice}$

\caption{Experience giving (EG) by advisor, $p_a$. $n^{v}$: no. of visits, $\varepsilon$: privacy budget, $s$: state, $a$: action, $B$: advice budget, $\eta$: LDP-noise, $d$: domain size}
\label{algo3}
\end{algorithm}
The fundamental mechanism to achieve $\varepsilon$-LDP is the randomized response (RR) \cite{wang2018locally}, a generalized version of which is \textit{Generalized Randomized Response (GRR)}, \cite{wang2018locally}. GRR is also described as a special Direct Encoding (DE) method 
and a generalization of k-randomized response \cite{wang2018locally}. 
 In GRR, given the domain size $d = |\mathcal{D}|$ and privacy budget, $\varepsilon$, the following perturbation probability ensures $\varepsilon$-LDP \cite{wang2018locally}.
\begin{equation}
    Pr\left[\mathcal{M}_{GRR}(x)=y\right]=\begin{cases}
\frac{e^{\varepsilon}}{d+e^{\varepsilon}-1} &\text{if $y=x$}\\
\frac{1}{d+e^{\varepsilon}-1} &\text{otherwise}
\end{cases}
\end{equation}
\begin{theorem}
    GRR satisfies $\varepsilon$-LDP.
\end{theorem}
\begin{proof}
    To satisfy $\varepsilon$-LDP, the ratio of the probabilities for $x, x' \in \mathcal{D}$ needs to be equal to $e^{\varepsilon}$. Here, we have
    \begin{equation}
        \frac{Pr\left[\mathcal{M}_{GRR}(x)=y\right]}{Pr\left[\mathcal{M}_{GRR}(x')=y\right]} = \frac{\frac{e^{\varepsilon}}{d+e^{\varepsilon}-1}}{\frac{1}{d+e^{\varepsilon}-1}} = e^{\varepsilon}
    \end{equation}
    which satisfies the condition of $\varepsilon$-LDP.
\end{proof}

In our setting, the advisors follow the GRR-based perturbation mechanism to achieve $\varepsilon$-LDP since GRR directly takes the original value as input into the perturbing step without the need for the encoding process. Following Definition \ref{def_LDP}, if we assume the set of all Q-values for every action in a particular state, $s_{p_i}$ as a dataset, $\mathpzc{D_a} = \{Q_{a_1},..., Q_{a_n}\}$, then the corresponding private (perturbed) dataset, $\mathpzc{D'_a} = \{Q'_{a_1},..., Q'_{a_n}\}$. The original Q-values $\{Q_{a_x}\}_{x=1}^{n}$ and private Q-values $\{Q'_{a_x}\}_{x=1}^{n}$ are linked by the privacy preservation mechanism $\mathcal{M}_{GRR}$. Here, $Q'_{a_x}$ depends only on $Q_{a_x}$; and not on any other Q-values $Q_{a_y}\text{ or } Q'_{a_y}$ for $y\neq x$. Therefore, this noninteractive framework can be given as
\begin{equation}
    Q'_{a_x} \leftarrow Q_{a_x} \text{ and } Q'_{a_x} \perp \{Q_{a_y}, Q'_{a_y}, y\neq x\}|Q_{a_x},
\end{equation}
where $\perp$ denotes the symbol of noninteractive relation. Algorithm~\ref{algo3} shows the pseudocodes (line $4$ to $7$) of the GRR mechanism for Q-value sharing. Given a privacy budget, $\varepsilon$, an original Q-value set $D_a$, the algorithm returns a perturbed Q-value set $D'_a$. Nonetheless, for any two neighboring Q-value sets of equal length (e.g., $\mathpzc{D_a} = \{Q_{a_1},..., Q_{a_n}\}$, $\mathpzc{D_b} = \{Q_{b_1},..., Q_{b_n}\}$, and $|\mathpzc{D}_a| = |\mathpzc{D_b}| = n$), the changes can occur for maximum $n$ positions. Therefore, the sensitivity of the mechanism is $n$ here. Specifically, the mechanism keeps a particular Q-value unchanged (i.e., $Q'_{a_x}\leftarrow Q_{a_x}$) with a probability, $p = \frac{e^{\varepsilon/n}}{d+e^{\varepsilon/n}-1}$ and perturbs it to a different random Q-value (i.e., $Q'_{a_x}\leftarrow \text{Uniform}(\mathcal{D}_a/Q_{a_x})$) with probability $q = \frac{1}{d+e^{\varepsilon/n}-1}$. 
\begin{proposition}
The proposed EG method satisfies $\varepsilon$-LDP.
\end{proposition}
\begin{proof}
The algorithm applies the GRR mechanism separately to each Q-value of a state learned by an advisor. If $\mathcal{M}_{i}(.)$ is applied on a particular Q-value, $x\in Q_{p_a}(s_{p_i}^t), \text{ where } |Q_{p_a}(s_{p_i}^t)| = n$ and the output is $y$, then 
 \begin{equation}
        \frac{Pr\left[\mathcal{M}_{i}(x)=y\right]}{Pr\left[\mathcal{M}_{i}(x')=y\right]} = \frac{\frac{e^{\varepsilon/n}}{d+e^{\varepsilon/n}-1}}{\frac{1}{d+e^{\varepsilon/n}-1}} = e^{\varepsilon/n}
    \end{equation}
\vspace{-2pt}
Therefore, following Theorem $2$, $\mathcal{M}_{i}(.)$ satisfies $\frac{\varepsilon}{n}$-LDP. Now, if we consider $\varepsilon_i = \frac{\varepsilon}{n}$, then we can combine $n$ subroutines (each satisfying $\varepsilon_i$-LDP independently) for $n$ number of Q-values by following the sequential property of $\varepsilon$-LDP given in Theorem $1$ for our EG algorithm and show that EG satisfies
$\sum_{i=1}^{n}\varepsilon_i = n\cdot \varepsilon_i = n \cdot \frac{\varepsilon}{n} = \varepsilon$-LDP.
\end{proof}
\textbf{Remark.} In our experimental setting, there are four Q-values for four corresponding actions (\textit{Left, Right, Up, Down}). Thus, $|\mathcal{D}_a| = 4$. Also, the maximum difference between two adjacent Q-value sets would be $4$. Hence, the applied GRR mechanism for each Q-value in a set satisfies $\frac{\varepsilon}{4}$-LDP, ensuring the overall EG method satisfies $4 \text{x} \frac{\varepsilon}{4} = \varepsilon$-LDP.
\vspace{-2pt}

\section{Experimental Analysis}

We implement our framework following a modified predator-prey domain \cite{le2017coordinated}. Next, we compare our results with two SOTA approaches: \textbf{AdhocTD} \cite{da2017simultaneously}, \textit{which proposes visit-based advising, but neither adopts DP nor incorporates weighted experience aggregation during ES}; and \textbf{DA-RL} \cite{ye2022differential}, \textit{which proposes a differential advising method but does not incorporate any neighbor zone concept and/or weighted experience aggregation technique to enable security and privacy-aware ES}.

Our environment is a $\mathpzc{H}\times \mathpzc{W}$ grid world with multiple agents and one goal. Agents have four actions to choose, \textit{Left, Right, Up, Down} to move from one cell to another. They can collect additional rewards upon visiting a freeway on the path to the goal. Nonetheless, grid obstacles can cause penalties upon encounter. Moreover, the agents get penalties if they hit any grid boundary. The positions of the agents, obstacles, and freeway are initialized randomly at the beginning of every episode. The game ends when all the agents reach the goal. Nonetheless, if the agents do not reach the goal within a predefined ($\text{grid size}\times 100$) steps, the environment is reset.
While we demonstrate our work in a grid world, it is also extendable to real-world domains with sensitive data. Table \ref{tab:parameter} lists the parameters we have used during our experiment.

We investigate the impact of the environment in three scenarios of different scales: (1) \textbf{small-scale:} $5\times5$ grid with $5$ agents, $1$ obstacle, and $1$ freeway, (2) \textbf{medium-scale:} $10\times 10$ grid with $10$ agents, $3$ obstacles, and $1$ freeway, and (3) \textbf{large-scale:} $30\times30$ grid with $20$ agents, $5$ obstacles, and $1$ freeway. We also consider varying percentages of attackers in each environment (e.g., no attacker, $20\%$ attackers, etc.). To evaluate and compare our results with AdhocTD \cite{da2017simultaneously} and DA-RL \cite{ye2022differential}, we use three popular metrics \cite{zhu2021q}: \textbf{Steps to goal (SG}), \textbf{Reward}, and \textbf{Time to goal (TG)}. SG is the average number of steps needs to reach the goal, Reward is the total average incentive earned, and TG is the total average learning time (in seconds) before reaching the goal. The experiments were conducted on a Lambda Tensorbook equipped with an $11$th Gen Intel(R) Core(TM) i$7$-$11800$H @$2.30$GHz CPU, RTX $3080$ Max-Q GPU, $64$ GB RAM, $2$ TB storage, Windows $10$ pro ($64$-bit) OS, Python $3.9.7$, and PyTorch $1.10.0+$cpu.
\begin{table}[!t]
\centering
\caption{Parameter value. $\alpha$: learning rate, $\epsilon$: exploration-exploitation probability, $\gamma$: discount factor, $B$: communication budget, $w$: aggregation factor, $\tau, \tau', \kappa$: predefined threshold, $\phi$: reward, $\varepsilon$: privacy budget.}
\label{tab:parameter}
\begin{adjustbox}{max width=\linewidth}
\begin{tabular}{c|c|c|c|c|c|c|c}
\toprule
Parameter &
$\alpha$                                                      & $\epsilon$                                                    & $\gamma$                                                      & \multicolumn{1}{c|}{$B^{tot}_{p_i}$}           & $B^{tot}_{p_a}$                          & $w$                                          & $\tau$   \\ 
\midrule
Value &
0.10                                                                           & 0.08                                                                           & 0.80                                                                           & 100,000                           & 10,000                                              & 0.85                                           & 100                       \\
\midrule
\multicolumn{1}{c|}{Parameter} &
\multicolumn{1}{c|}{$\phi_{\mathpzc{G}}$} & \multicolumn{1}{c|}{$\phi_{\mathpzc{F}}$} & \multicolumn{1}{c|}{$\phi_{\mathpzc{O}}$} & $\phi_{\mathpzc{W}}$ & \multicolumn{1}{l|}{$\varepsilon$} & \multicolumn{1}{c|}{$\kappa$} & $\tau'$  \\ 
\midrule
\multicolumn{1}{c|}{Value}  & 
\multicolumn{1}{c|}{10.0}                                                      & \multicolumn{1}{l|}{0.50}                                                      & \multicolumn{1}{c|}{-1.50}                                                     & -0.50                                                     & \multicolumn{1}{l|}{1.0}                            & \multicolumn{1}{l|}{0.1}                      & 100,000   \\
\bottomrule
\end{tabular}
\end{adjustbox}
\vspace{-15pt}
\end{table}
\subsection{Trajectory Analysis}
\begin{figure}[!ht] 
\vspace{-10pt}
\centerline{\includegraphics[width=\linewidth]{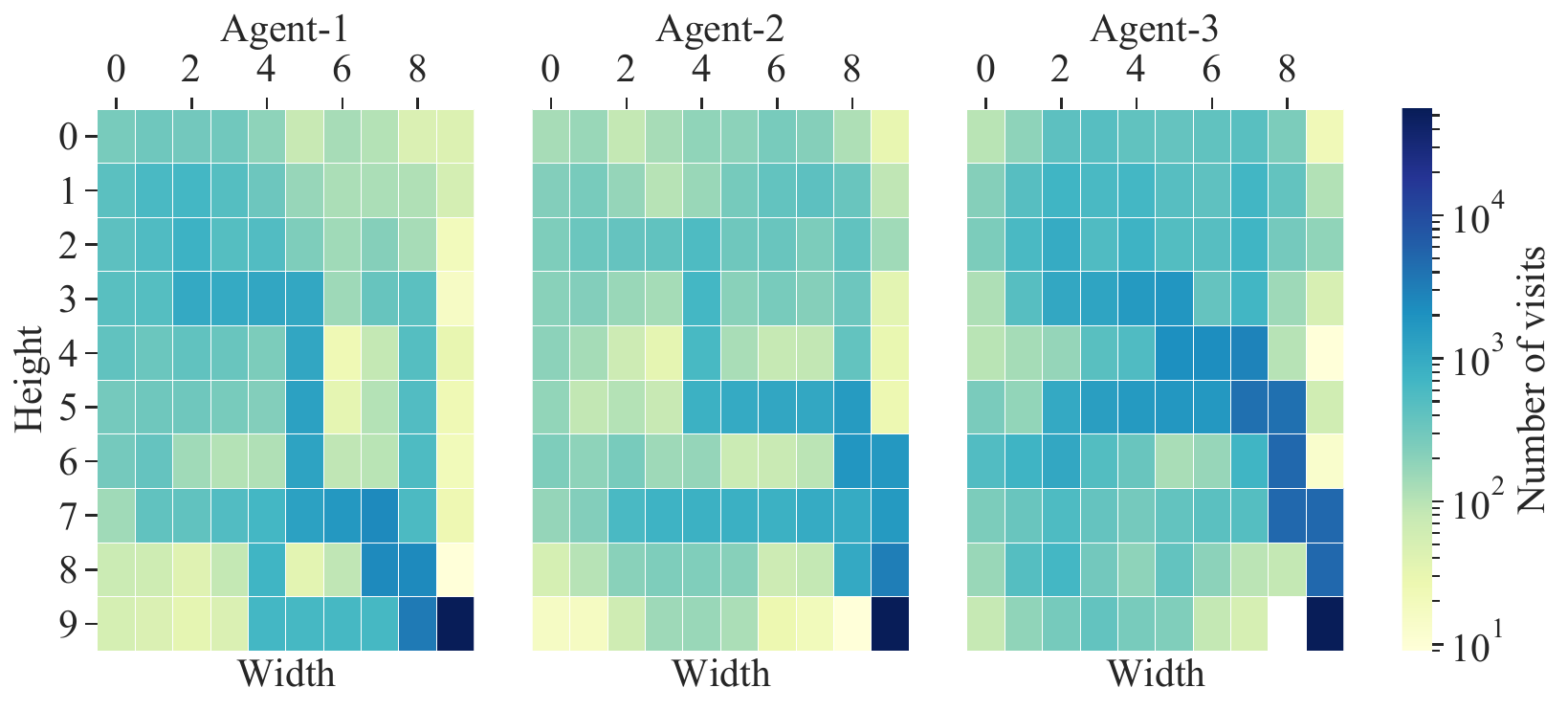}}
    \caption{Visiting trajectory of the agents.} 
    \label{fig:trajectory}
    \vspace{-10pt}
\end{figure}

We perform a trajectory analysis of the agents. The result is illustrated in Figure~\ref{fig:trajectory}. The cells with darker colors have been visited more frequently than the cells with lighter colors. It can be inferred that all of the agents have visited the cells that are closer to the goal more frequently as compared to the cells that are far distant from the goal. Another interesting fact is that in most cases, the agents have a lower tendency to visit the boundary cells, which in turn provides evidence that the agents have learned to avoid hitting the grid boundaries and getting penalties. 

\begin{figure*}[!t]
\centerline{\includegraphics[width=\linewidth]{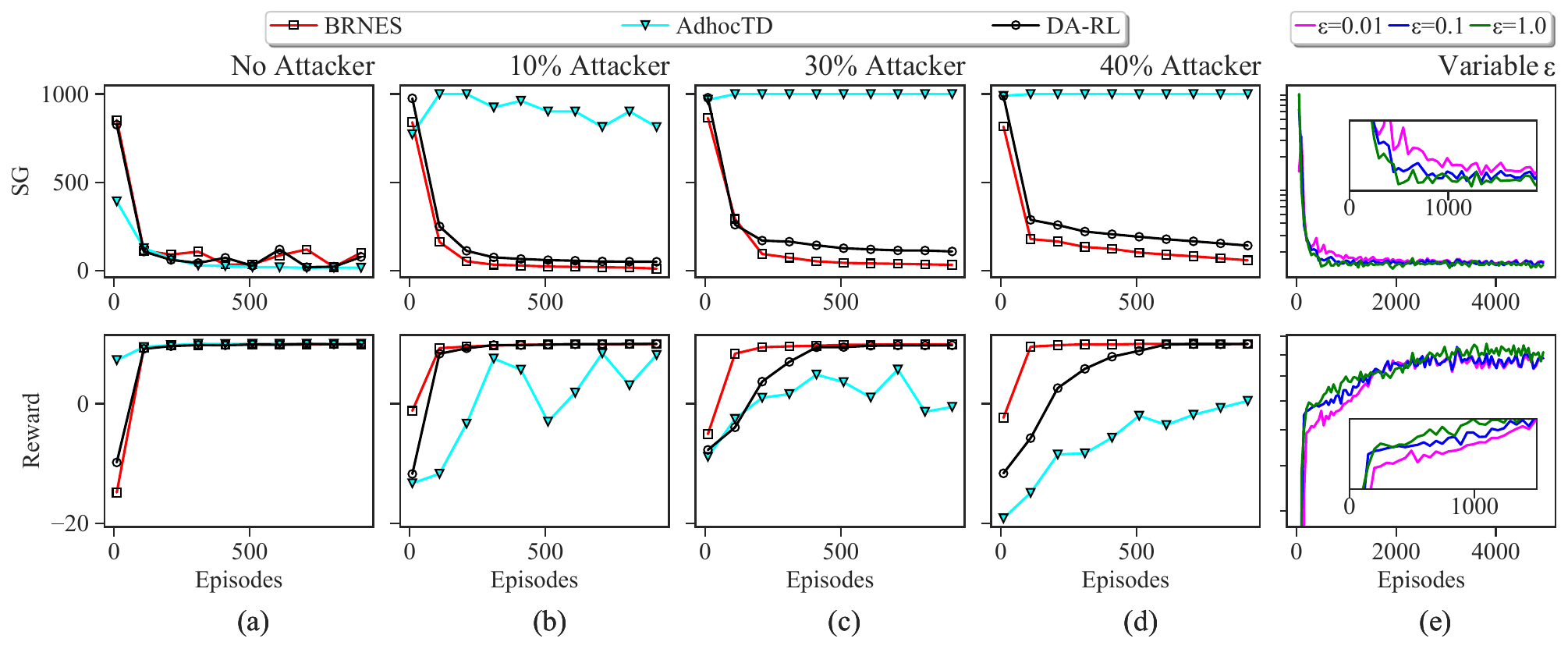}}
\vspace{-10pt}
    \caption{Steps to goal (SG) and Reward comparison for $1000$ episodes among AdhocTD \cite{da2017simultaneously}, DA-RL\cite{ye2022differential} and Our (BRNES) framework. Environment ($\mathbb{E}$) feature: [$(\mathpzc{H}\times \mathpzc{W}): (10 \times 10), \mathpzc{G}: 1, \mathpzc{F}: 1, \mathpzc{O}: 3, \lvert \mathpzc{N} \rvert = 10$]. (a) Baseline scenario (No Attacker), (b)-(d) Multiple Attackers, (e) Variable privacy budget.} 
    \label{fig:SGRewardVarEps}
    
\end{figure*}

\subsection{Steps to Goal (SG) and Reward Analysis} Figure~\ref{fig:SGRewardVarEps}a-\ref{fig:SGRewardVarEps}d reflects the average SG values and corresponding rewards of our framework (BRNES), AdhocTD \cite{da2017simultaneously}, and DA-RL \cite{ye2022differential} in a medium-scale environment under \textit{no attacker} and \textit{multiple attackers} scenarios. Lower SG values indicate that the agents reach the goal more quickly, and vice versa. When there is no attacker (Figure~ \ref{fig:SGRewardVarEps}a), all of the frameworks have lower SG values from early episodes (i.e., $<200$ episodes). Particularly, AdhocTD \cite{da2017simultaneously} exhibits the most stable performance in \textit{no attacker} cases (Figure~\ref{fig:SGRewardVarEps}a). This is mostly because it does not incorporate any DP noise and thus, incurs zero privacy cost. Nonetheless, despite having some privacy overhead, BRNES continues to closely follow AdhocTD \cite{da2017simultaneously} and outperforms DA-RL \cite{ye2022differential} for \textit{no attacker} case. In contrast, as soon as Byzantine advisors appear, the SG values of AdhocTD \cite{da2017simultaneously} rapidly grow. The more the concentration of the attacker among the agents, the more the SG values. This can be observed in Figure~\ref{fig:SGRewardVarEps}b-\ref{fig:SGRewardVarEps}d, which illustrates that \textit{BRNES outperforms both AdhocTD \cite{da2017simultaneously} and DA-RL \cite{ye2022differential} in multiple attacker scenarios}. Reward graphs, underneath the corresponding SG graphs, exhibit similar results. Specifically, in Figure~\ref{fig:SGRewardVarEps}b-\ref{fig:SGRewardVarEps}d where AdhocTD \cite{da2017simultaneously} and DA-RL \cite{ye2022differential} obtained optimal reward after approximately $300, 400, \text{and } 600$ episodes, BRNES continues to indicate significant improvement in learning by obtaining optimal rewards in earlier episodes.
\begin{figure}[!t]
\vspace{-10pt}
\centerline{\includegraphics[width=\linewidth]{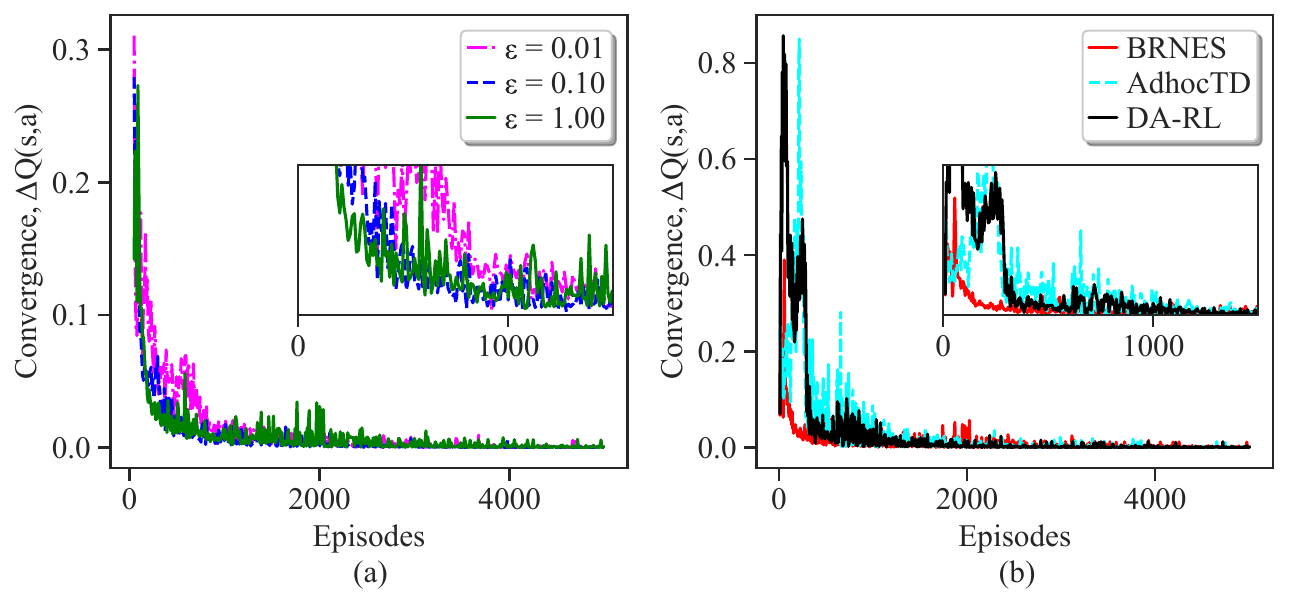}}
\vspace{-10pt}
    \caption{(a) Convergence is faster when privacy is low (i.e., large $\varepsilon$), (a) BRNES converges faster than AdhocTD \cite{da2017simultaneously} and DA-RL\cite{ye2022differential}. Both (a) and (b) are in a medium-scale environment with $30\%$ attackers. } 
    \vspace{-15pt}
    \label{fig:convergence}
\end{figure}

\subsection{Impact of Privacy Budget}
We evaluate our framework for multiple values of privacy budget, $\varepsilon$. As shown in Figure~\ref{fig:SGRewardVarEps}e, BRNES performs better for higher $\varepsilon$ (i.e., low privacy regime) in terms of both SG and Reward. Also, Figure~\ref{fig:convergence}a depicts that the convergence happens faster for $\varepsilon = 1.00$ compared to $\varepsilon = 0.01$, which also supports the privacy-utility tradeoff scenario of DP, i.e., higher privacy, lower utility, and vice versa.
\vspace{-2pt}
\subsection{Convergence Analysis} Convergence analysis under adversarial presence is depicted in Figure~\ref{fig:convergence}b. It is evaluated based on the average values of the $\Delta Q (s^t, a^t)$ for all $\Delta Q = Q(s^{t+1}, a^{t+1}) - Q(s^t, a^t)$. The key idea is to show the Q-values are converging into the optimal Q value ($Q^*$). For simplicity, we only present the deterministic case, in which $Q(s^{t+1}, a^{t+1})$ converges to $Q^*(s, a)$. Therefore, if the average of $\Delta Q(s, a)$ goes to zero, BRNES can be considered stable. From Figure~\ref{fig:convergence}b, it can be seen that $\Delta Q (s, a)$ gradually goes to zero. Nonetheless, while AdhocTD \cite{da2017simultaneously} and DA-RL \cite{ye2022differential} are converging after around $900$ and $400$ episodes respectively, BRNES converges faster (i.e., in $<200$ episodes). 

\subsection{Time to Goal (TG) Analysis} 
\begin{figure}[!b]
\vspace{-10pt}
\centerline{\includegraphics[width=\linewidth]{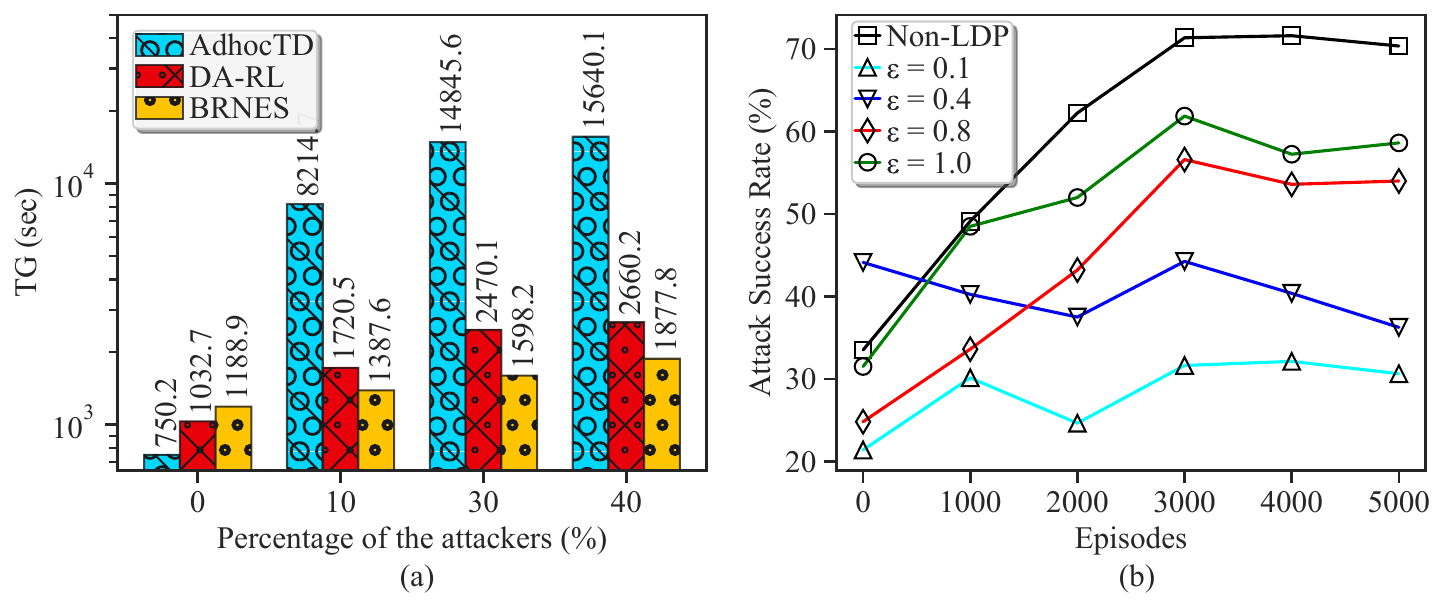}}
\vspace{-10pt}
    \caption{(a) TG comparison under adversarial presence, (b) Inference attack success rate (\%) is the lowest when privacy is the highest (i.e., $\varepsilon=0.1$).} 
    \vspace{-12pt}
\label{fig:TGandInference}
\end{figure}

TG value comparison is presented in Figure~\ref{fig:TGandInference}a and Table \ref{tab:TGcomparison}. BRNES requires the lowest time for the agents to reach the goal, except for $0\%$ attackers cases since it deploys LDP-noise to enable private experience sharing, which leads to noisy Q-values. In addition to this privacy cost, the neighbor zone selection and weighted aggregation technique also incur some computational overhead. Nonetheless, this overhead becomes insignificant for BRNES as compared to other frameworks under adversarial presence (Figure~\ref{fig:TGandInference}a and Table~\ref{tab:TGcomparison}). Particularly, for $40\%$ attacker case in a medium scale-environment, BRNES is $(15640.1/1877.8)\approx 8.32$x faster than AdhocTD \cite{da2017simultaneously}, and $(2660.2/1877.8)\approx 1.41$x faster than DA-RL \cite{ye2022differential} in terms of TG value metric.
\vspace{-2pt}
\subsection{Protection from Inference Attacks} To empirically evaluate the effectiveness of our LDP-driven BRNES framework against inference attacks, we compare multiple $\varepsilon$ scenarios with a baseline Non-LDP scenario
We observe how accurately and quickly an attacker could infer the movement of an advisor by performing repeated advising requests. The results, as shown in Figure~\ref{fig:TGandInference}b, demonstrate that the Non-LDP baseline scenario allows an attacker to achieve a success rate of approximately over $70\%$ within $3000$ episodes. However, as we adopt LDP through our proposed framework and increase privacy protection (i.e., decrease $\varepsilon$), the attack success rate decreases significantly. 

\begin{table}[!t]
\caption{Experimental result for Time to goal (TG).}
  \label{tab:TGcomparison}
  \vspace{-5pt}
   \begin{adjustbox}{max width=\linewidth}
\begin{tabular}{c|c|c|c|c}
\toprule
\multirow{2}{*}{\begin{tabular}[c]{@{}c@{}}Environment \\Type\end{tabular}}   & \multirow{2}{*}{\begin{tabular}[c]{@{}c@{}}Attacker \\(\%Agent)\end{tabular}} & \multirow{2}{*}{\begin{tabular}[c]{@{}c@{}}AdhocTD \cite{da2017simultaneously} \\(TG (sec)\end{tabular}}  & \multirow{2}{*}{\begin{tabular}[c]{@{}c@{}}DA-RL \cite{ye2022differential} \\(TG (sec)\end{tabular}}     & \multirow{2}{*}{\begin{tabular}[c]{@{}c@{}}BRNES \\(TG (sec)\end{tabular}}     \\ 

                               &                               &       &  &     \\ 
\midrule
\multirow{3}{*}{{}small-scale} & 0\%                           & \textbf{91.3}     & 699.7    & 552.4     \\ 
\cline{2-5}
                               & 20\%                          & 3125.1   & 776.5    & \textbf{584.9}     \\ 
\cline{2-5}
                               & 40\%                          & 4170.2   & 970.1    & \textbf{754.2}     \\ 
\hline
\multirow{3}{*}{{}medium-scale} & 0\%                           & \textbf{750.2}    & 1032.7   & 1188.9    \\ 
\cline{2-5}
                               & 30\%                          & 14845.5   & 2470.1   & \textbf{1598.2}    \\ 
\cline{2-5}
                               & 40\%                          & 15640.1  & 2660.2   & \textbf{1877.8}    \\ 
\hline
\multirow{3}{*}{{}large-scale} & 0\%                           & \textbf{45487.5}  & 71479.4  & 61693.8   \\ 
\cline{2-5}
                               & 30\%                          & 164852.8 & 95172.9  & \textbf{73740.8}   \\ 
\cline{2-5}
                               & 40\%                          & 245425.7 & 146295.7 & \textbf{103787.3} \\
   \bottomrule
\end{tabular}
\end{adjustbox}
\vspace{-15pt}
\end{table}

\section{Conclusion}
In this study, to mitigate the adversarial impact during experience sharing in CMARL, we propose a novel framework, BRNES, that strategically incorporates neighbors' experiences for effective and faster convergence. 
Our framework outperforms the SOTA approaches in terms of steps to goal (SG), reward, and time to goal (TG) while achieving $\varepsilon$-LDP to mitigate inference attacks. \textbf{Specifically, our framework achieves $8.32\text{x}$ faster TG than a non-private framework, AdhocTD \cite{da2017simultaneously}, and $1.41\text{x}$ faster TG than a private framework, DA-RL \cite{ye2022differential} in a medium-scale environment under adversarial presence.}

Several interesting extensions emerge for future privacy and security research in MARL, including analyzing adversarial activity in fully cooperative or competitive and mixed cooperative-competitive environments. Our framework could be extended to a more dynamic environment, where agents receive new tasks when they complete their current tasks.

\bibliographystyle{IEEEtran}
\bibliography{ref}

\end{document}